\documentclass[%
 reprint,
 amsmath,amssymb,
 aps,
]{revtex4-2}

\usepackage{graphicx}
\usepackage{color}
\usepackage{dcolumn}
\usepackage{bm}
\usepackage[utf8]{inputenc}
\usepackage[T1]{fontenc}
\usepackage{mathptmx}
\usepackage{etoolbox}
\usepackage{color}
\usepackage{dcolumn}
\usepackage{bm}
\usepackage[mathlines]{lineno}
\usepackage{comment}

\begin{document}

\preprint{APS/123-QED}

\title{Spin Hall effect in van der Waals ferromagnet Fe$_{5}$GeTe$_{2}$}

\author{T.~Ohta\textsuperscript{1}}
\email{ohta@meso.phys.sci.osaka-u.ac.jp}
\author{Y.~Samukawa\textsuperscript{1}}
\author{N.~Jiang\textsuperscript{1,2,3}}
\author{Y.~Niimi\textsuperscript{1,2,3}}
\author{K.~Yamagami\textsuperscript{4}}
\author{Y.~Okada\textsuperscript{5}}
\author{Y.~Otani\textsuperscript{6,7}}
\author{K.~Kondou\textsuperscript{3,7}}
\email{{kondou.kouta.otri@osaka-u.ac.jp}}

\affiliation{\textsuperscript{1}Department of Physics, Graduate School of Science, Osaka University, Osaka 560-0043, Japan}
\affiliation{\textsuperscript{2}Center for Spintronics Research Network, Osaka University, Osaka 560-8531, Japan}
\affiliation{\textsuperscript{3}Institute for Open and Transdisciplinary Research Initiatives, Osaka University, Osaka 565-0871, Japan}
\affiliation{\textsuperscript{4}Japan Synchrotron Radiation Research Institute,Hyogo 679-5198, Japan}
\affiliation{\textsuperscript{5}Okinawa Institute of Science and Technology Graduate University, Okinawa 904-0495, Japan}
\affiliation{\textsuperscript{6}Institute for Solid State Physics, The University of Tokyo, Chiba 277-8581, Japan}
\affiliation{\textsuperscript{7}RIKEN Center for Emergent Matter Science (CEMS), Saitama 351-0198, Japan}





\date{\today}

\begin{abstract}
We investigate the spin Hall effect (SHE) in a van der Waals (vdW) ferromagnet Fe$_{5}$GeTe$_{2}$ (FGT) with a Curie temperature $T_{\rm C}$ of 310~K utilizing the spin-torque ferromagnetic resonance method. In synchronization with the emergence of the ferromagnetic phase resulting in the anomalous Hall effect (AHE), a noticeable enhancement in the SHE was observed below $T_{\rm C}$. On the other hand, the SHE shows a different temperature dependence from the AHE: the effective spin Hall conductivity is clearly enhanced with decreasing temperature unlike the anomalous Hall conductivity, reflecting the variation of band-structure accompanied by the complicated magnetic ordering of the FGT. The results provide a deep understanding of the SHE in magnetic materials to open a new route for novel functionalities in vdW materials-based spintronic devices. 
\end{abstract}




\maketitle


\section{\label{sec:level1}INTRODUCTION}

The spin Hall effect (SHE) was theoretically proposed by Dyakonov and Perel in 1971~\cite{dyakonov_PLA_1971} and experimentally demonstrated by Kato $et$ $al$. in 2004~\cite{kato_science_2004}. It enables the interconversion of charge and spin currents in the transverse direction~\cite{sinova_revmodphys_2015}, applied as a spin-charge interconversion method for devices such as non-volatile magnetic memory and energy harvesting devices from light, sound, and heat to charge current~\cite{otani_natphys_2017}. The SHE has mainly been investigated in nonmagnetic heavy metals with strong spin-orbit coupling, but it can be observed even in magnetic materials~\cite{wei_natcommun_2012, miao_prl_2013, taniguchi_prappl_2015, iihama_natele_2018, bose_pra_2018, das_prb_20217, gibbons_pra_2018, seki_prb_2019, wang_natnanotech_2019, kimata_nature_2019, kondou_natcommun_2021, you_natcommun_2021}. So far, there have been many theoretical and experimental studies on the SHE unique to magnetic materials: for example, spin-anomalous Hall effect (AHE) in ferromagnetic metals (FMs)~\cite{taniguchi_prappl_2015, iihama_natele_2018, das_prb_20217, gibbons_pra_2018, seki_prb_2019}, anomalous spin-orbit torque in FMs~\cite{bose_pra_2018, wang_natnanotech_2019}, and magnetic SHE in topological antiferromagnets~\cite{kimata_nature_2019, kondou_natcommun_2021, you_natcommun_2021}. For these phenomena, the spin polarization direction due to SHE can be controlled by manipulating the magnetic moment. It would enable us to realize novel functions such as perpendicular magnetization switching~\cite{legrand_physrevappl_2015, kong_natcommun_2019} and high-speed efficient magnetization switching~\cite{legrand_physrevappl_2015} in magnetic devices. Thus, understanding the SHE in magnetic materials is favorable for future spintronic research.
Recently, unconventional SHEs with unique spin polarization vectors have been reported in low-symmetry materials~\cite{liu_natnanotechnol_2021, kao_natmater_2022, yu_natnanotechnol_2014, macneill_natphys_2017, song_natmater_2020}. For example, spin-orbit torque (SOT) due to out-of-plane spin polarization has been observed in some transition metal dichalcogenides (TMDCs)~\cite{macneill_natphys_2017, song_natmater_2020}. In vdW FMs, combining with heavy metals has demonstrated the SOT-induced magnetization switching~\cite{ostwal_advmater_2020, alghamdi_nanolett_2019}. However, the spin Hall effect in vdW FMs has never been investigated. 
Here, we focus on a metallic vdW ferromagnet, Fe$_{5}$GeTe$_{2}$ (FGT), with a Curie temperature $T_{\rm C}$ of 310~K~\cite{stahl_allgchem_2019, may_prm_2019, ohta_apex_2020, ohta_apl_2023, yamagami_prb_2022} to study the magnetic state dependent SHE in vdW FMs. Relatively high $T_{\rm C}$ and the temperature-dependent unique magnetic anisotropy make this material a fascinating candidate for spintronic devices based on atomic-layer materials. Owing to its high $T_{\rm C}$, FGT enables us to investigate the AHE and the SHE above and below $T_{\rm C}$. In this work, we measured the SHE in FGT in a wide temperature range employing the spin-torque ferromagnetic resonance (ST-FMR) method. We found that the conversion efficiency from charge to spin current, i.e., the effective spin Hall angle, is strongly enhanced with the emergence of ferromagnetism in FGT. It takes a constant value of 0.2, well below $T_{\rm C}$. On the other hand, the AHE shows a different temperature dependence from the SHE: it increases with decreasing temperature but starts to fall below 120~K, indicating that an ordering of Fe1 site in FGT plays an essential role in reducing the AHE. The magnitude of the effective spin Hall conductivity becomes comparable to those of typical spin Hall materials such as Pt and W at low temperatures. Our findings not only shed light on the potential of a vdW magnet as a spin-Hall material but also provide a deeper understanding of the SHE in magnetic materials.

\section{\label{sec:level2}EXPERIMENTAL SETUP}

\subsection{\label{subsec:level1}Magnetic and electrical transport properties of van der Waals ferromagnet FGT }
Figure 1(a) shows the crystal structure of FGT, which is a hexagonal lattice with a space group of $R\bar{3}m$ No. 166. The unit cell comprises three slabs stacked along the $c$-axis, each with Fe and Ge atoms sandwiched by Te layers. The unit cell contains three Fe sites (Fe1, Fe2, and Fe3): the Fe2 and Fe3 sites are ordered below $T_{\rm C}$, while the Fe1 site is ordered below 120~K~\cite{may_prm_2019, ohta_apex_2020, ohta_apl_2023}. 
In order to confirm the basic physical properties of our FGT, we have first checked both thin film (Figs.~1(b), 1(c) and 1(e)) and bulk (Fig.~1(d)) properties. We show the temperature dependence of magnetization $M$ for one of our bulk single crystals in Fig.~1(d)~(~\cite{yamagami_prb_2022}, see also Supplemental Materials~\cite{ohta_prm_ref}). The temperature dependences of the longitudinal resistivity $\rho_{xx}$ (left axis) and the anomalous Hall resistivity $\rho_{\rm{AH}}$ (right axis) of a 70~nm thick device (see Fig. 1(b)) are also shown in Fig.~1(e). $\rho_{\rm{AH}}$ is extracted by extrapolating the normal Hall component at high magnetic fields to zero (see Fig.~1(c))~\cite{ohta_prm_ref}). $M$ and $\rho_{\rm{AH}}$ emerge below $T_{\rm C}$ ($\approx310$~K) and increase with temperature down to 150~K. In the lower temperatures region, however, $\rho_{\rm{AH}}$ is strongly suppressed below 120~K, while $M$ along the $c$-axis is slightly changed below 150~K. Such complex behavior is considered as follows. In general, $\rho_{\rm{AH}} = R_{\rm{s}}\textit M_{z}$, where $R_{\rm{s}}$ is the anomalous Hall coefficient, and $M_{z}$ is the magnetization along the $c$-axis. For FGT, $R_{\rm{s}}$ is known to be almost constant above 150~K, showing that $\rho_{\rm{AH}}$ is roughly proportional to $M_{z}$ in this temperature region. Below 150~K (see the caption of Fig. 1), on the other hand, $\rho_{\rm{AH}}$ is not simply proportional to $M_{z}$, indicating that $R_{\rm{s}}$ changes~\cite{may_prm_2019}. It is suggested that two ordered states are related to the temperature dependence of $\rho_{\rm{AH}}$ and $M_{z}$ at low temperatures: the periodic charge ordering below 150~K~\cite{wu_prb_2021} and the magnetic ordering at the Fe1-site below 120~K~\cite{may_prm_2019, ohta_apex_2020}. The maximum shape of $M$ around 150~K corresponds to the charge ordering attributed to the Fermi surface nesting~\cite{wu_prb_2021}, which is thought to be strongly correlated to the change in magnetic anisotropy~\cite{wu_prb_2021, tang_natelectron_2023}.  The emergence of new electron-pocket at around 120~K have been reported~\cite{wu_prb_2021}. This might be related to the drastic changes in longitudinal and Hall resistivities and the sign change of the Hall coefficient at 120~K~\cite{may_prm_2019, wu_prb_2021}.

\begin{figure}
\begin{center}
\includegraphics[width=9cm]{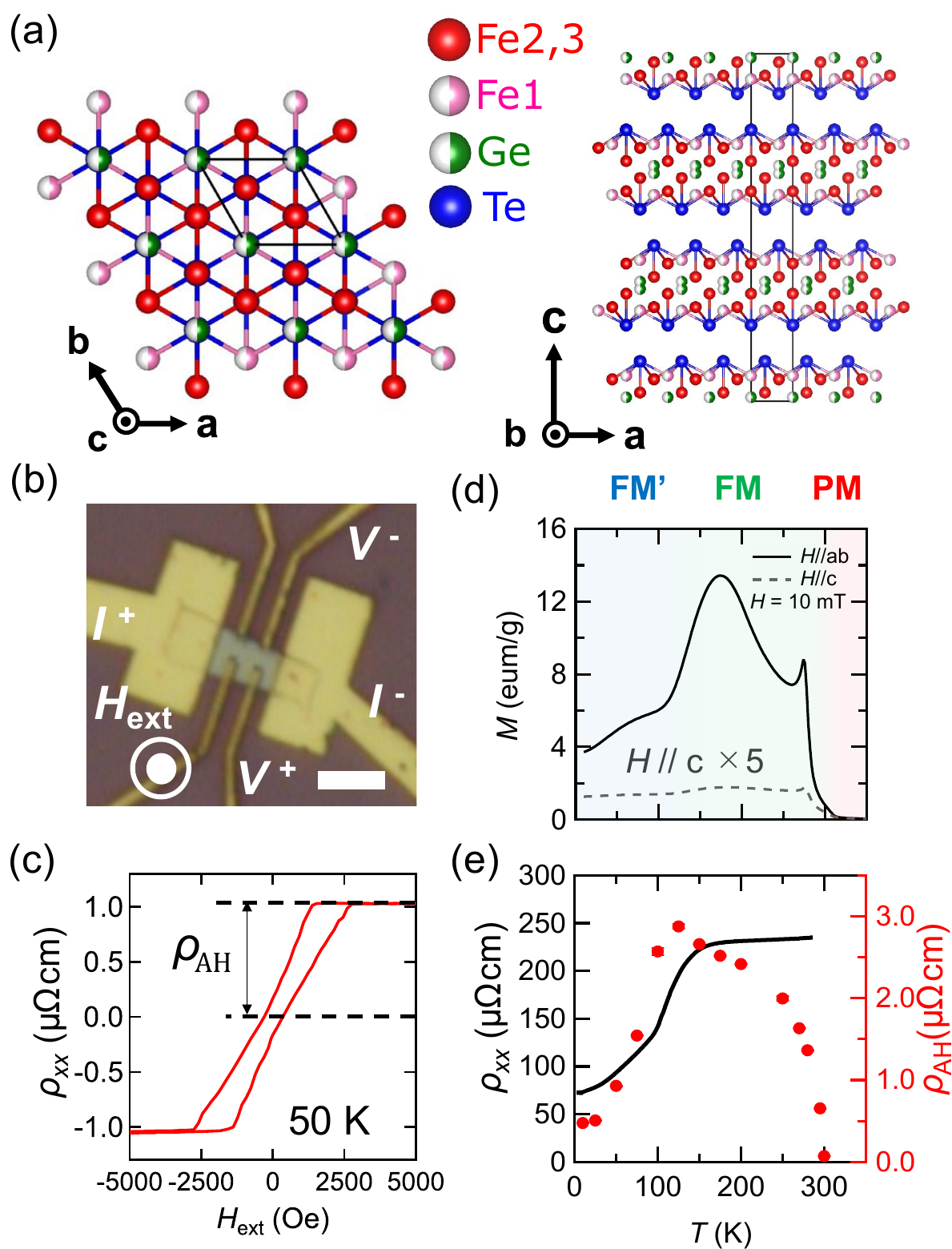}

\caption{(a)~Crystal structure of FGT depicted by VESTA~\cite{vesta}. The top and side views of the crystal are illustrated on the left and right, respectively. The gray rectangle in the right panel represents a unit cell. Fe1 and Ge sites are 50\% occupied. (b) Optical microscope image of FGT device for the Hall measurement. A current is applied to the in-plane direction and a magnetic field is applied to the out-of-plane direction as shown in the image. The scale bar corresponds to 10 $\rm{\mu}$m. (c) Anomalous Hall effect at 50 K. The definition of the anomalous Hall resistivity is also indicated. (d) Temperature dependence of magnetization measurement of the bulk sample. Black and gray lines show the in-plane ($H//ab$) and the out-of-plane ($H//c$) magnetization. $H//c$ is multiplied by 5 for the clarity. The magnetic state of FGT changes depending on the temperature region. The FM and PM regions correspond to the ferromagnetic and paramagnetic phases, respectively, while the FM’ region is defined as the region where the in-plane magnetization decreases with decreasing temperature. (e) Temperature dependence of the longitudinal resistivity (left axis) and the anomalous Hall resistivity (right axis) obtained from the thin film sample.}
\label{figure1}
\end{center}
\end{figure}

\subsection{\label{subsec:level2}Fabrication of FGT/Cu/Ni-Fe trilayer device}
To investigate the SHE in FGT with the magnetic phase transition, we fabricated FGT(70 nm)/Cu(5 nm)/Ni$_{81}$Fe$_{19}$(Ni-Fe;10~nm) trilayer devices for spin-torque ferromagnetic resonance (ST-FMR) measurements as shown in Fig. 2(a) Ref~\cite{ohta_prm_ref}. In such relatively thicker FGT films, the in-plane magnetic anisotropy is more dominant than the out-of-plane component~\cite{tang_natelectron_2023, zhang_prb_2020, ohta_prm_ref}. We first carried out the mechanical exfoliation using Scotch tapes in a glovebox filled with 99.9999\% Ar gas to prevent the oxidation of FGT flakes. The exfoliated FGT flakes with a thickness of about 70~nm were transferred onto a SiO$_{2}$/Si substrate. We then coated polymethyl-methacrylate resist on the substrate and patterned for Cu/Ni-Fe deposition with electron beam lithography. After the lithography, the resist was developed in the glovebox, followed by the Cu and Ni-Fe deposition onto the FGT flakes to fabricate the trilayer microstrips in a vacuum chamber next to the glovebox with the base pressure of $5\times10^{-6}$~Pa. We note that before the deposition of Cu and Ni-Fe, Ar milling was performed to remove the residual resist and any possibly oxidized layers of FGT. After the first deposition and lift-off process, we coated the resist again and patterned the waveguide structure electrodes with lithography. Similarly, Ti (5~nm) and Au (100~nm) were deposited as the contact waveguide. After device fabrications, all the devices were capped with SiO$_{2}$ (5~nm) by rf-sputtering to prevent deterioration of the devices. After all electrical measurements, the film thickness was determined using commercial atomic force microscopy. 

\begin{figure}
\begin{center}
\includegraphics[width=8.5cm]{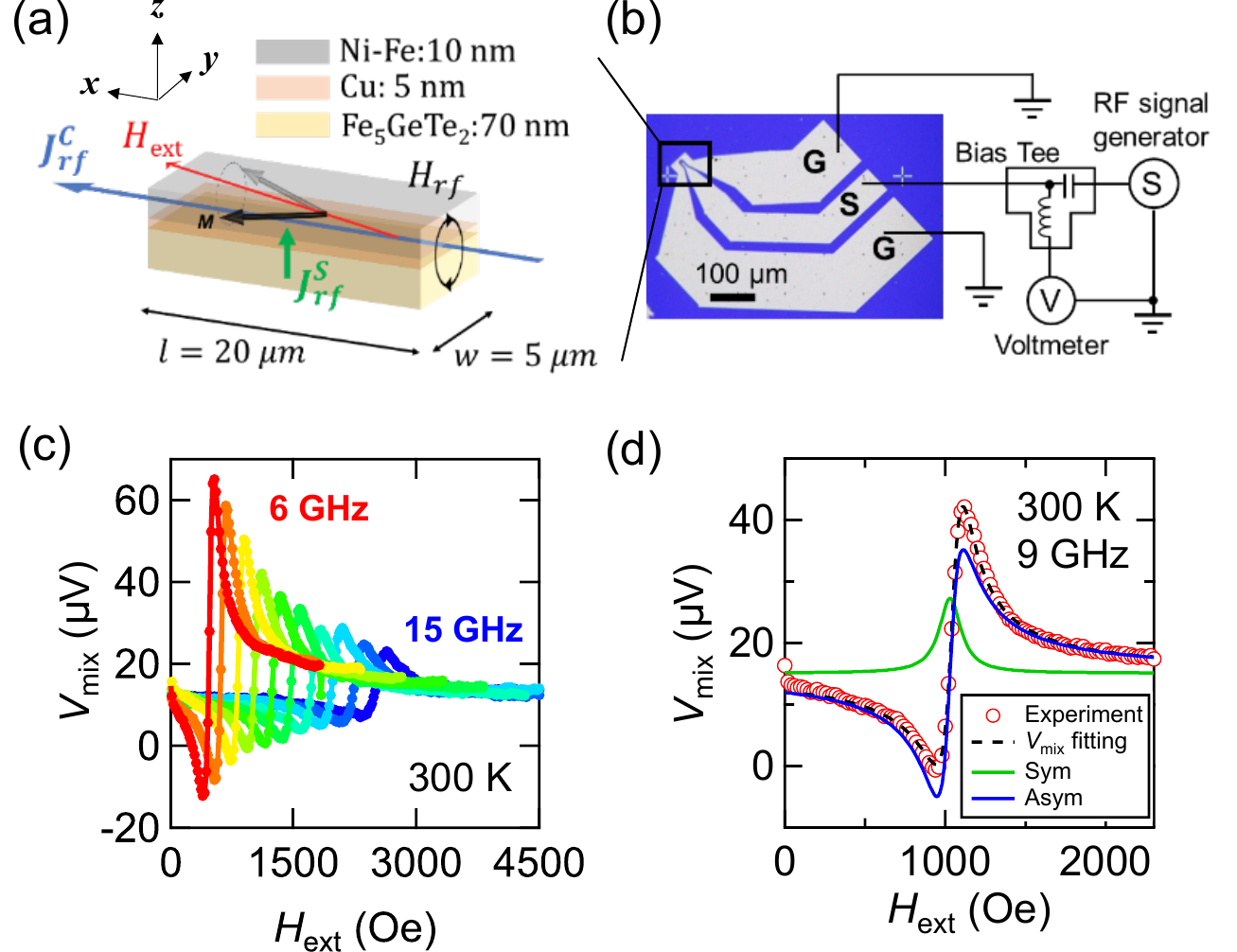}
\caption{(a)~Schematic of ST-FMR device consisting of FGT(70)/Cu(5)/Ni-Fe(10). Magnetization $M$ precession of Ni-Fe, external magnetic field $H\rm{_{ext}}$, rf current $J\rm{^{C}_{rf}}$, rf magnetic field $H\rm{_{rf}}$, and spin current generated by the SHE in FGT $J\rm{^{S}_{rf}}$ are illustrated as arrows. (b) Schematic of the circuit of the ST-FMR measurement setup. (c) FMR spectra obtained with FGT(70)/Cu(5)/Ni-Fe(10) device. The frequency range is from 6 to 15 GHz. (d) FMR spectrum at $T = 300$~K and $f = 9$~GHz. The black dotted line is the best fit with Eq. (1). The green and blue lines show the symmetric and antisymmetric components of the black dotted curve, respectively. The yellow line shows the contribution from the Ni-Fe single layer.}
\label{figure2}
\end{center}
\end{figure} 

\subsection{\label{subsec:level3}Spin-torque ferromagnetic resonance method}
ST-FMR measurements were performed in the temperature range from 25~K to 350~K. The size of the microstrip of the trilayer device is approximately 5-$\rm{\mu}$m-wide, 20-$\rm{\mu}$m-long with Au/Ti ground-signal-ground electrodes. Figure 2(b) shows the schematic diagram of the ST-FMR setup. The Cu layer was inserted to avoid any direct exchange coupling between the Ni-Fe and FGT layers. In this trilayer device, the rf charge-current ($J\rm{^{C}_{rf}}$) in the FGT and Cu layers induces an Oersted field $H\rm{_{rf}}$. This field exerts the out-of-plane torques on the magnetization of Ni-Fe, resulting in the ferromagnetic resonance. At the same time, $J\rm{^{C}_{rf}}$ induces a transverse spin-current ($J\rm{^{S}_{rf}}$) via the SHE in FGT. The spin current is injected into the adjacent Ni-Fe layer through the Cu layer, where the spin current exerts the in-plane torque i.e., damping-like torque on the magnetization of Ni-Fe.
The magnetization precession induces an oscillating anisotropic magnetoresistance (AMR) in Ni-Fe. The rf charge-current and the oscillating sample resistance with the same frequency produce a rectified dc voltage ($V\rm{_{dc}}$) across the sample in the FMR state. The spectral analysis of the $V\rm{_{dc}}$ signal yields the symmetric and asymmetric components, respectively, corresponding to the in-plane and out-of-plane torque contributions by considering the phase of these torques differs by $\pi/2$. 

An in-plane external field $H\rm{_{ext}}$ with a fixed angle of $45^{\circ}$ from the trilayer wire (Fig. 2(a)) was swept between $\pm4~\rm{k}{Oe}$. Since the microwave skin depth is much larger than the Ni-Fe thickness, the current distribution in Ni-Fe is spatially uniform. The Oersted field can, therefore, be calculated from the sum of the current density from the Cu and FGT layers. The width of the conducting channels is much larger than the total thickness of ($t\rm{_{Cu}}$ + $t\rm{_{FGT}}$). Thus, the device can be treated as an infinitely wide conducting plate. The Oersted field determined by the Ampere’s low in our devices is thus given by $H\rm{_{rf}}$ = $J\rm{_{C,rf}^{FGT}}\frac{\textit t_{FGT}}{2}$ + $J\rm{_{C,rf}^{Cu}}\frac{\textit t\rm{_{Cu}}}{2}$, where $J\rm{_{C,rf}^{FGT} }$ and $J\rm{_{C,rf}^{Cu}}$ represent the charge current densities in the FGT and Cu layers, respectively~\cite{yang_apl_2020}. The partial current flowing in the FGT layer can be calculated by considering the resistance ratio between FGT and Cu layers. This calculation enables us to estimate the magnitude of the Oersted fields from the FGT layer. All the ST-FMR measurements were performed with an input rf power of 10~dBm.

\section{\label{sec:level3}EXPERIMENTAL RESULTS AND DISCUSSIONS}
We show ST-FMR spectra obtained for an FGT/Cu/Ni-Fe trilayer device with the applied rf current frequency $f$ ranging from 6 to 15~GHz at 300~K in Fig. 2(c). We also show ST-FMR spectra at different temperatures in Ref~\cite{ohta_prm_ref}. Increasing the rf frequency shifts the resonant magnetic field $H\rm{_0}$ to the higher magnetic field. Figure 2(d) is a typical ST-FMR spectrum at $f$ = 9~GHz fitted with the following equation~\cite{liu_prl_2011}: 

        \[
        V_{\rm{mix}} = 
        - \frac{1}{4} 
        \frac{dR}{d\theta} 
        \frac{\gamma I_{\rm{rf}} \, {\rm{cos}}\theta}
        {\Delta H 2 \pi \left( \frac{df}{dH} \right) \bigg|_{H_{\rm{ext}} = H_{\rm{0}}}} 
        \]
    \begin{equation}
        \times \left[ V_{\rm S} F_{\rm S}(H_{\rm{ext}}) + V_{\rm A} F_{\rm A}(H_{\rm{ext}}) \right],
    \end{equation}

\noindent where $R$ is the resistance of the trilayer microstrip line, $\theta$ is the angle between $M$ of Ni-Fe and the microstrip line, $\gamma$ is the gyromagnetic ratio, $I_{\rm{rf}}$ is the rf current, $\Delta H$ is the FMR linewidth~\cite{ohta_prm_ref}, $V_{\rm S} = \hbar J_{\rm{rf}}^{\rm C} / (2e\mu_{\rm{0}} M_{\rm{s}} t_{\rm{NiFe}})$ is the amplitude of symmetric voltage, and  $V_{\rm A} = H_{\rm{rf}} [1 + (4\pi M_{\rm{eff}}/H_{\rm{ext}})]^{1/2}$ is that of antisymmetric voltage. Here, $\hbar$ is the reduced Plank constant, $e$ is the elementary charge, $\mu_{\rm{0}}$ is the permeability in a vacuum. $M_{\rm{s}}$, $t_{\rm{NiFe}}$ and $4\pi M_{\rm{eff}}$ are the saturation magnetization, the thickness, and the demagnetization field of Ni-Fe, respectively. The green and blue curves in Fig. 2(d) are the Lorentzian and anti-Lorentzian components for the resonant field $H_{\rm{0}}$, i.e., $F_{\rm S}(H_{\rm{ext}}) = \Delta H^{2}/[\Delta H^{2} + (H_{\rm{ext}} - H_{\rm{0}})^{2}]$ and $F_{\rm A} (H_{\rm{ext}}) = \Delta H(H_{\rm{ext}} - H_{\rm{0}})/[\Delta H^{2} + (H_{\rm{ext}} - H_{\rm{0}})^{2}]$, respectively. A symmetric voltage $V_{\rm S}$ due to spin pumping into the FGT is negligibly smaller than the observed $V_{\rm S}$ shown in Fig.~2(d)~\cite{ohta_prm_ref}. In addition, the contribution from the self-induced torque of the Ni-Fe(10) layer is evaluated and subtracted from the total amplitude (as discussed in Ref.~\cite{ohta_prm_ref}). We have checked the effective saturation magnetization of Ni-Fe by using the in-plane Kittel formula $f = (\gamma/2\pi) [H_{\rm{0}} (H_{\rm{0}}  + 4\pi M_{\rm{eff}} )]^{1/2}$~\cite{ohta_prm_ref} and confirmed that $4\pi M_{\rm{eff}}$ of the Ni-Fe layer is about $8.75$~kG (see also Fig. S6), consistent with the previous work~\cite{ounajela_jdp_1988}.

\begin{figure}
\begin{center}
\includegraphics[width=8.0cm]{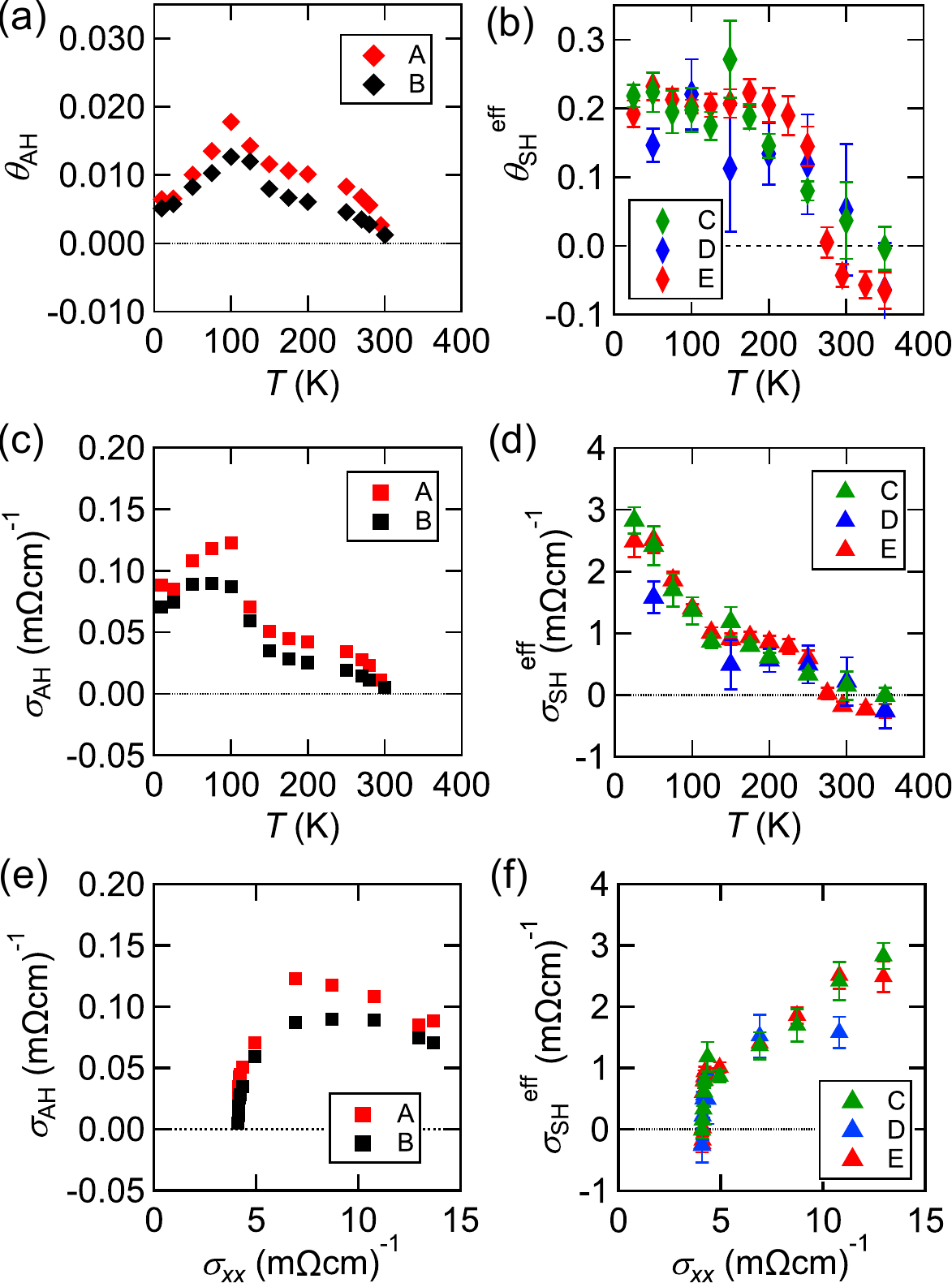}
\caption{(a) Temperature dependence of the anomalous Hall angle for FGT device A (70 nm) and device B (60 nm). The same data as shown in Fig. 1(e) are used for device A. (b) Temperature dependence of the effective spin Hall angle for FGT devices C-E (60-70 nm). The same data as shown in Fig. 2 are used for device C. (c), (d) Temperature dependence of the anomalous Hall conductivity (c) and the effective spin Hall conductivity (d). (e), (f) The anomalous Hall conductivity (e) and the effective spin Hall conductivity (f) as a function of the longitudinal conductivity. The dotted lines in all figures indicate zero value.}
\label{figure3}
\end{center}
\end{figure}

To discuss the SHE in our FGT, we evaluated the temperature dependence of the effective spin Hall angle $\theta^{\rm{eff}}_{\rm{SH}} (=J\rm{^{S}_{rf}} / J\rm{^{C}_{rf}})$ utilizing the ST-FMR method. Since the symmetric ($V_{\rm S}$) and antisymmetric ($V_{\rm{A}}$) voltages in the spectrum include $J\rm{^{S}_{rf}}$ and $J\rm{^{C}_{rf}}$, the ratio of $J\rm{^{S}_{rf}}$ and $J\rm{^{C}_{rf}}$, i.e., $\theta^{\rm{eff}}_{\rm{SH}}$ can be estimated by using the ratio of $V_{\rm S}$ and $V_{\rm A}$ as shown in following equation:

\begin{align}
    \theta^{\rm{eff}}_{\rm{SH}} &= \frac{J\rm{^{S}_{rf}}}{J\rm{^{C}_{rf}}} \notag \\
    &= \frac{V_{\rm S}}{V_{\rm A}} \frac{e\mu_{\rm{0}} M_{\rm{s}} t_{\rm{NiFe}}}{\hbar} (\frac{\rho_{\rm{FGT}}}{\rho_{\rm{Cu}}}t_{\rm{Cu}} + t_{\rm{FGT}}) (1 + \sqrt{\frac{4\pi M_{\rm{eff}}}{H_{\rm{ext}}}}).
\end{align}
In Eq. (2), we have considered a partial current flowing into the FGT layer from the resistance ratio between FGT and Cu by measuring the temperature-dependent $\rho_{\rm{FGT}}$ and $\rho_{\rm{Cu}}$ for individual FGT and Cu films with the same thicknesses as shown in Ref~\cite{ohta_prm_ref}.  Note that we have taken the average of positive and negative field sides of the spectrum to calculate $\theta^{\rm{eff}}_{\rm{SH}}$ in order to exclude the thermal contribution. The obtained $\theta^{\rm{eff}}_{\rm{SH}}$ for three FGT devices are plotted as a function of temperature in Fig. 3(b).

As a reference, we show the temperature dependence of the anomalous Hall angle, $\theta_{\rm{AH}} (=\rho_{\rm{AH}} / \rho_{xx})$, for two FGT samples in Fig. 3(a). The $\theta_{\rm{AH}}$ emerges below $T_{\rm{C}}$, takes a maximum at 100~K, and decreases at lower temperatures. As shown in Fig. 1(e), the increase of $\theta_{\rm{AH}}$ below $T_{\rm{C}}$ reflects the emergence of $M$, while the sudden decrease of $\theta_{\rm{AH}}$ below 100~K does not follow the $M$. It suggests that the quadratic contribution of $\rho_{xx}$ to $\rho_{\rm{AH}}$ is dominant, as $\rho_{\rm{AH}}$ can be expressed as $a(\rho_{xx}) + b(\rho_{xx})^{2}$ (where $a$ and $b$ are prefactors)~\cite{tian_prl_2009, Nagaosa_rmp_2010}. Figure 3(c) shows the temperature dependence of anomalous Hall conductivity (AHC) $\sigma_{\rm{AH}}$. Since $\sigma_{\rm{AH}}$ = $\rho_{\rm{AH}}$ / ($\rho_{\rm{xx}}^2$ + $\rho_{\rm{AH}}^2$) $\approx$ $\rho_{\rm{AH}}$ / $\rho_{\rm{xx}}^2 = a/\rho_{\rm{xx}} + b$, $\sigma_{\rm{AH}}$ is almost constant below 100~K, suggesting the significance of the quadratic contribution. We can also confirm this trend from the $\rho_{xx}$ dependence of $\theta_{\rm{AH}}$ in Ref~\cite{ohta_prm_ref}. This behavior is consistent with recent experimental results on the AHE in FGT~\cite{deng_nanolett_2022} and the temperature dependence of $R_{\rm S}$~\cite{stahl_allgchem_2019}. 

Figures 3(b) and 3(d) show the temperature dependence of $\theta^{\rm{eff}}_{\rm{SH}}$ and effective spin Hall conductivity (SHC) $\sigma^{\rm{eff}}_{\rm{SH}}$, respectively. The $\theta^{\rm{eff}}_{\rm{SH}}$ takes almost zero or a slightly negative value above $T_{\rm C}$, while it changes the sign below $T_{\rm C}$ and reaches about $0.2$ well below $T_{\rm C}$. Such tendency is reproducible for three devices, as shown in Fig. 3(c). It indicates that the SHE in FGT is sensitive to the magnetic ordering of the Fe2 and Fe3 sites as in the case of the AHE. At the moment, the detailed mechanism of the enhancement of $\theta^{\rm{eff}}_{\rm{SH}}$ is an open question, but it is clear that $\theta^{\rm{eff}}_{\rm{SH}}$ increases according to the ferromagnetic transition near $T_{\rm{C}}$~\cite{ohta_prm_ref}.

Recently, there have been several reports on the magnetization direction-dependent SHE in ferromagnets~\cite{wang_natnanotech_2019, chuang_prr_2020, yagmur_prb_2021, davidson_pla_2020}. If the magnetization of FGT affects the spin polarization direction, it may appear on the spectrum as a change in the torque component acting on the magnetization of Ni-Fe. As mentioned in Sec.\ref{subsec:level2}, we have selected relatively thick FGT samples where the magnetization of FGT mainly points to the in-plane direction ($x$-$y$ plane in Fig. 2(a)), i.e., $\bm{m_{x,y}}$, under $H_{\rm{ext}}$ of more than $500$~Oe~\cite{tang_natelectron_2023}.We have also confirmed that the in-plane magnetization is dominant in our 70 nm devices from the Hall measurements~\cite{ohta_prm_ref}. In this situation, if $\bm{s_{y}}$, the spin polarization direction owing to the SHE in FGT processes with respect to $\bm{m_{x,y}}$, the conduction electron spin can have some components along the $z$-axis, i.e., $\bm{s_{z}}$~\cite{wang_natnanotech_2019, beak_natmatter_2018}. Since the magnetization in the NiFe layer $M_{\rm{NiFe}}$ is also in the in-plane direction, the damping-like-torque due to $\bm{s_{z}}$ gives rise to the out-of-plane torques, i.e., $\bm{M_{\rm{NiFe}}\times(s_{z}\times M_{\rm{NiFe}})}$, which should affect the amplitude of $V_{\rm A}$. However, there were no significant differences in the amplitude of $V_{\rm A}$ below and above $T_{\rm C}$ in our devices~\cite{ohta_prm_ref}. Since the thickness of the FGT layer is much larger than that of the Ni-Fe layer, the Oersted field is expected to dominate $V_{\rm A}$. Such a large $V_{\rm A}$ can mask a small contribution from unconventional torque. Therefore, we do not rule out the presence of unconventional torque contributions in this structure.

To elucidate this enhancement of $\theta^{\rm{eff}}_{\rm{SH}}$ according to the ferromagnetic transition, the magnetization direction dependence of $\theta^{\rm{eff}}_{\rm{SH}}$ could provide some clues. For example, if there exists a new type of SHE induced by magnetization onset, it should respond to the magnetization reversal of the FGT. Thus, by utilizing thinner FGT films with stronger perpendicular magnetization and switching the magnetization direction of FGT by $H_{\rm{ext}}$ in thinner films, more detailed information on magnetization direction-dependent SHE in the FGT could be obtained.  Measurements with thin FGT would also improve our sensitivity to detect the unconventional torque by reducing the dominant Oersted-field torque contribution. This is an important direction for the future work. 

Below the ordering temperature of Fe1 sites, on the other hand, the effective SHC shown in Fig. 3(d) is clearly enhanced unlike the AHC shown in Fig. 3(c). Such a difference can also be seen in the longitudinal conductivity $\sigma_{\rm{xx}}$ dependence of the AHC and SHC (Figs. 3(e)-3(f)). One possible scenario for the different behavior between the AHC and SHC at lower temperatures is the contribution of the band-structure. According to a recent report by X. Wu \textit{et al}.~\cite{wu_prb_2021}, the modulation of band-structure below the ordering temperature of Fe1 sites ($\sim120$~K) gradually proceeds with decreasing temperature. The difference between the AHC and SHC could be explained by differences in the symmetry of the Berry curvature and spin-Berry curvature~\cite{qu_jpsj2021}. In the paper, they showed that the deviation between the Berry curvature and the spin Berry curvature appears in a typical ferromagnet with the symmetry reduction of the spin Berry curvature. This originates from the different spin direction of the Bloch state at the band-crossing point and becomes more significant if we take into account the anomalous velocity term. Therefore, the spin polarized density functional theory calculation would provide some clues to elucidate the intrinsic SHE in FGT.

\begin{figure}
\begin{center}
\includegraphics[width=8cm]{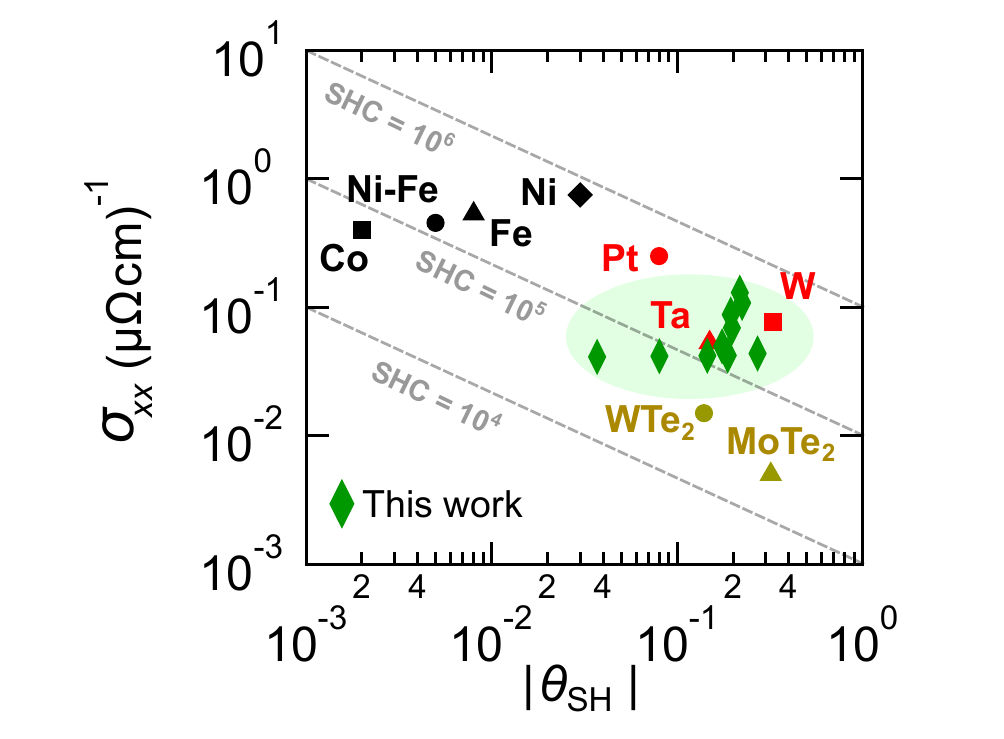}
\caption{Relation between electrical conductivity $\sigma_{xx}$ and the effective spin Hall angle |$\theta^{\rm{eff}}_{\rm{SH}}$| for various spin Hall materials. We plot the results of device C as the green symbols and also green shaded area. Red, black, and yellow symbols are the results of 4$d$ and 5$d$ transition metals~\cite{liu_prl_2011, pai_apl_2012, liu_science_2012}, 3$d$ ferromagnets~\cite{omori_prb_2019}, and 2D materials~\cite{song_natmater_2020, zhao_prr_2020}. The grey dotted line represents a fixed effective spin Hall conductivity, which is written as the product of |$\theta^{\rm{eff}}_{\rm{SH}}$| and $\sigma_{xx}$.The dotted line used $(\rm{\Omega m)}^{-1}$ as a unit.}
\label{figure4}
\end{center}
\end{figure}

Before closing the section, we evaluated the figure of merit for FGT as a spin Hall material. For this purpose, in Fig. 4, we show the relation between $\sigma_{xx}$ and the absolute value of the effective spin Hall angle |$\theta^{\rm{eff}}_{\rm{SH}}$| for various spin Hall materials such as 4$d$ and 5$d$ transition metals, 3$d$ ferromagnetic metals~\cite{omori_prb_2019}, and nonmagnetic vdW materials. If $|\sigma^{\rm{eff}}_{\rm{SH}}|$ (gray dotted lines in Fig. 4) is larger, the spin current is generated more efficiently. The green symbols show the obtained results in this work. We found that $|\sigma^{\rm{eff}}_{\rm{SH}}|$ for FGT is almost comparable to that of typical spin Hall materials such as Pt and W~\cite{liu_prl_2011, pai_apl_2012, liu_science_2012}. Compared to nonmagnetic vdW materials such as $\rm{WTe_{2}}$ and $\rm{MoTe_{2}}$~\cite{song_natmater_2020, zhao_prr_2020}, $|\sigma^{\rm{eff}}_{\rm{SH}}|$ for FGT takes slightly higher values. Importantly, the $|\sigma^{\rm{eff}}_{\rm{SH}}|$ can vary one order of magnitude by changing the temperature.

\section{\label{sec:level4}CONCLUSIONS}
In conclusion, we have demonstrated ST-FMR measurements with FGT(70~nm)/Cu(5~nm)/Ni-Fe(10~nm) trilayer devices and evaluated the magnetic phase-dependent SHE in FGT. 
The temperature dependence of the effective spin Hall angle $\theta^{\rm{eff}}_{\rm{SH}}$ shows almost zero or slightly negative value above $T_{\rm C}$, while the sign of $\theta^{\rm{eff}}_{\rm{SH}}$ changes from negative to positive across $T_{\rm C}$ and shows an almost constant value well below $T_{\rm C}$. The observed enhancement of $\theta^{\rm{eff}}_{\rm{SH}}$ below $T_{\rm C}$ is very similar to that of $\theta_{\rm{AH}}$, indicating that the SHE in this material is sensitive to the magnetic ordering. At lower temperatures, on the other hand, $\theta^{\rm{eff}}_{\rm{SH}}$ shows completely different temperature dependence from $\theta_{\rm{AH}}$, suggesting the existence of the spin Hall effect accompanied by the complicated magnetic ordering of the FGT. We have also revealed that $\sigma^{\rm{eff}}_{\rm{SH}}$ for FGT is almost comparable to that of typical spin Hall materials such as Pt and W and enhanced around one order of magnitude below $T_{\rm C}$ by changing temperature. These findings provide not only the new insight into the SHE in magnetic materials but also the potential application for magnetic-phase-controlled electronic devices based on vdW magnetic materials.

\begin{acknowledgments}
We thank T. Kawada, T. Taniguchi, K. Ueda, S. Iwakiri, F. Bonell for the fruitful discussion. The lattice structure of FGT was visualized using VESTA~\cite{vesta}. This work was supported by JSPS KAKENHI (Grant Nos. JP20H02557, JP21J20442, JP22H04481, and JP23H00257), JST FOREST (Grant No. JPMJFR2134), and RIKEN-Osaka University Science $\&$ Technology Hub Collaborative Research Program.
\end{acknowledgments}

\nocite{*}


\end{document}